\documentstyle[12pt,aaspp4]{article}

\lefthead{Sembach et al.}
\righthead{Modeling the WIM}

\begin{document}

\newcommand{\degr}{$^{\circ}$}
\newcommand{\kms}{\,km\,s$^{-1}$}     
\newcommand{\Ha}{H$\alpha$}

\title{Modeling the Warm Ionized Interstellar Medium and Its
Impact on Elemental Abundance Studies}

\author{Kenneth R. Sembach \& J. Christopher Howk}
\affil{Department of Physics \& Astronomy, The Johns Hopkins University,
3400 N. Charles St., Baltimore, MD  21218 \nl
{\it sembach@pha.jhu.edu, howk@pha.jhu.edu}}

\author{Robert S.I. Ryans \& Francis P. Keenan}
\affil{Department of Pure \& Applied Physics, The Queen's University of 
Belfast, \nl
Belfast, BT7 1NN, Northern Ireland \nl
{\it R.Ryans@qub.ac.uk, F.Keenan@qub.ac.uk}}

\begin{abstract}
We present model calculations of ionization fractions for elements in
the warm ($T\sim10^4$\,K), low-density photoionized interstellar
medium (WIM) of the Milky Way.  We model the WIM as a
combination of overlapping low-excitation \ion{H}{2} regions having
$n(H^+)/n(H) \gtrsim 0.8$.  Our adopted standard model incorporates an
intrinsic elemental abundance pattern similar to that found for warm
neutral clouds in the Galaxy and includes the effects of
interstellar dust grains.  The
radiation field is characterized by an ionizing spectrum
of a star with $T_{eff} \approx 35,000$\,K and an
ionization parameter $\log (q) \approx -4.0$.  The emergent emission
line strengths are in agreement with the
observed ratios of [\ion{S}{2}]/H$\alpha$, [\ion{N}{2}]/H$\alpha$,
[\ion{S}{2}]/[\ion{N}{2}], [\ion{O}{1}]/H$\alpha$,
[\ion{O}{3}]/H$\alpha$, and \ion{He}{1}/H$\alpha$ in the Galactic
WIM.  Although the forbidden emission-line intensities
depend strongly upon the input model parameters, the ionization
fractions of the 20 elements studied in this work are
robust over a wide range of physical conditions considered in the 
models.  These ionization
fractions have direct relevance to absorption-line determinations of
the elemental abundances in the warm neutral and ionized gases
in the Milky Way and other late-type galaxies.  We
demonstrate a method for estimating the WIM contributions to the 
observed column densities of singly and doubly ionized atoms used to
derive abundances in the warm neutral gas. We apply this approach to study the
gas-phase abundances of the warm interstellar clouds toward the halo
star HD\,93521.

\end{abstract}


\section{Introduction}

In elemental abundance studies of the warm, neutral interstellar medium
(WNM), it 
is generally assumed that ionization corrections for photoionized gas are 
negligible, or at least modest.  In practice, this assumption usually
takes the form $N(X)/N(H) \approx N(X^i)/N(H^0)$, where $X^i$ is the 
dominant ionization stage of element ``$X$'' in the neutral gas, even though
$X^i$ may also be found in ionized regions (see Savage \& Sembach 1996).
Ignoring ionized gas contributions to $N(X^i)$ is often justified when
studying cold atomic and molecular clouds, but this simplification
introduces 
additional uncertainties into studies of the interstellar medium (ISM)
in the immediate vicinity of
early-type stars or along extended, low-density sight lines.  
In these latter cases, ionized  material
can represent a significant fraction of the total amount of interstellar
gas observed and can contribute to the column densities of 
singly and doubly ionized atoms.  

The warm (T$\sim$10$^4$ K), ionized interstellar medium (WIM) outside
classical \ion{H}{2} regions is a fundamental gas-phase constituent of
the Milky Way and other late-type spiral galaxies.  In the Milky Way,
the WIM is detected through faint optical emission-line radiation
and pulsar dispersion measures, which together reveal that the ionized
gas has an exponential scale height of $\sim1$ kpc, a volume filling factor of
$\gtrsim 20$\%, and an average electron density of $\sim0.1$ cm$^{-3}$ 
(Reynolds 1993).  Though not as easily observed as the \ion{H}{1} layer 
revealed through 21\,cm emission, the WIM has a mean mass surface density 
($\sim 1.6$ M$_\odot$ pc$^{-2}$) nearly
one-third that of the \ion{H}{1} distribution (Reynolds 1993; 
Dickey \& Lockman 1990).  Pervasive warm ionized gas distributions have been 
observed in other spiral galaxies besides the Milky Way to varying extents
(Dettmar 1990; Rand, Kulkarni, \& Hester 1990, 1992; Rand 1996;
Hoopes, Walterbos, \& Rand 1999).  In all cases it seems clear that
while there may be a variety of ionization sources and mechanisms
contributing to the WIM emission, photoionization by OB-stars is the
only source of energy sufficient to maintain the bulk ionization
properties of the WIM (which in the Milky Way requires $\ga 10^{-4}$ erg
cm$^{-2}$ s$^{-1}$, or $\sim15\%$ of the ionizing radiation 
output from OB-stars). 

There are two recent observational motivations for modeling the WIM and its
impact on elemental abundance studies of the ISM.
First, there is an increasing database of 
emission line measurements of the WIM.  In 
particular, the Wisconsin H$\alpha$ Mapper (WHAM, Reynolds et al. 1998b) is 
providing wide sky coverage of H$\alpha$ and other optical emission lines
in the diffuse gas.
Second, the Far Ultraviolet
Spectroscopic Explorer (FUSE) will soon produce high spectral
resolution absorption-line measurements in the 905--1195\,\AA\ bandpass, 
which contains many diagnostics of photoionized gases.  Together with 
absorption-line data from the Hubble Space Telescope Imaging Spectrograph 
(HST/STIS), FUSE will enable comprehensive studies of the elemental abundances 
in the WNM and WIM.

In this paper, we estimate ionization fractions for the WIM based upon
existing knowledge of its properties.  These ionization fractions can
be used to derive appropriate corrections for ionized gas
contributions to the measured amounts of singly and doubly ionized
species found in both the WNM and WIM.  We begin in \S2 by
summarizing the current observational constraints on the WIM.  In \S3
we describe our model of the WIM, its boundary conditions, the
parameter space explored, and the sensitivity of the predicted
emission line ratios and ionization fractions to the input
parameters. In \S4 we provide a short description of the applicability
of the model results to absorption-line studies of elemental
abundances in the ISM.  Section 5 contains a brief discussion of these
results and a few concluding remarks.  Ionization fractions for
radially-averaged sight lines through low density \ion{H}{2} regions
can be found in Appendix A. 

\section{Emission-Line Properties of the Warm Ionized Medium}

The primary optical emission lines used to characterize the WIM are
H$\alpha$\,$\lambda$6563, [\ion{S}{2}]\,$\lambda$6716,
[\ion{N}{2}]\,$\lambda$6583, [\ion{O}{1}]\,$\lambda$6300,
[\ion{O}{3}]\,$\lambda$5007, and \ion{He}{1}\,$\lambda$5876.  We
compile a set of measurements of these species appearing in the
literature in Table~1, where we list the line intensities normalized
to the observed H$\alpha$ intensity.  Compared to classical \ion{H}{2}
regions, the WIM has high [\ion{N}{2}]/H$\alpha$ and
[\ion{S}{2}]/H$\alpha$ and low [\ion{O}{3}]/H$\alpha$ ratios
(Reynolds 1985b; Osterbrock 1989).  A subset of the listed ratios has
been modeled and discussed by Domg\"{o}rgen \& Mathis (1994), who found
that dilute stellar radiation leaking through the ISM provides an
adequate source of ionization for the WIM.  This paradigm is supported
by models of the radiative transfer of ionizing photons
through the Galaxy subject to the presence of interstellar clouds (Miller
\& Cox 1993; Dove \& Shull 1994; Bland-Hawthorn \& Maloney 1999).  
Additional secondary ionizing sources may be present; in at least one case 
(Ogden \& Reynolds 1985), the emission from a filament within the
diffuse background emission appears to be due to collisional ionization
by a weak
shock.  However, the large [\ion{N}{2}]/H$\alpha$ ratio is atypical of the
more general value of 0.4--0.6 for the emission observed in other
directions (see Table~1).

The emission-line ratios for gas in the Local and Perseus spiral arms change
significantly as a function of the observed H$\alpha$ intensity
(Haffner, Reynolds, \& Tufte 1999).  Ratios of [\ion{S}{2}]/H$\alpha$
and [\ion{N}{2}]/H$\alpha$ in excess of unity are seen in some
directions where I$_{H\alpha} < 0.5$ R.  The ranges of [\ion{S}{2}]/H$\alpha$
and  [\ion{N}{2}]/H$\alpha$ listed in
Table~1 for the Perseus arm encompass a majority of the observed values
and are appropriate for heights $z \la 1.0$ kpc
above the Perseus arm.  Haffner et al. (1999) argue that the
[\ion{N}{2}]/H$\alpha$ ratio probes the temperature of the gas, while
the [\ion{S}{2}]/[\ion{N}{2}] ratio traces the ionization state of
sulfur.  (The [\ion{S}{2}]/H$\alpha$ ratio is therefore a probe of
both physical properties.)  They interpret the large,
systematic increase in [\ion{N}{2}]/H$\alpha$ with height above
the plane as an indicator of increasing temperature, with an as yet
unidentified source of heating at large distances from the Galactic
midplane.  Extragalactic background radiation is an unlikely source,
since the ionizing flux out to about 100 kpc from the Galactic disk is
dominated by light from the Galaxy (Maloney \& Bland-Hawthorn 1999).
Interestingly, even though the [\ion{S}{2}]/H$\alpha$ and
[\ion{N}{2}]/H$\alpha$ ratios change by more than a factor of three
with height above the plane in the Perseus arm, the ratio
[\ion{S}{2}]/[\ion{N}{2}]$\, \approx 0.5 - 0.65$ is relatively
constant and shows virtually no dependence on height above the
Galactic midplane.  Thus, in the interpretation presented by Haffner
et al., the increase in temperature with height above the plane is not
matched by corresponding changes in the ionization state of the gas.

To date, the strongest constraints on the ionization of the WIM come
from [\ion{O}{1}]\,$\lambda$6300, [\ion{O}{3}]\,$\lambda$5007, and
\ion{He}{1}\,$\lambda$5876 emission-line observations (see
Domg\"{o}rgen \& Mathis 1994).  A complete set of optical emission-line
measurements is not yet available even for a single direction; we make
the simplifying assumption that the range in values and upper limits
listed in Table~1 are representative of the WIM as a whole.  However,
we also recognize that a range of physical conditions is possible and
allow the model parameters to vary accordingly.  Additional emission-line data 
in the near future will hopefully lead to a better description of the 
intensities of the less well-studied species.

\section{Model Description}

To study the physical conditions of the Galactic WIM, we use the
ionization equilibrium code CLOUDY (v90.04; described by
Ferland et al. 1998 and Ferland 1996) to derive the temperature and
ionization structure of spherically-symmetric nebulae surrounding
single ionizing sources.  We model the WIM as a combination of
low-density, low-excitation \ion{H}{2} regions.  Our goal is to estimate
the ionization fractions, $x(X^i) \equiv N(X^i) / N(X)$, of several
ionization stages, $i$, of the most important elements, $X$, for
studying the gas-phase abundances of the Galactic ISM.  In doing so we
endeavor to match the observed emission-line observations of the WIM
as closely as possible.  

\subsection{Approach and Assumptions}

Following Domg\"{o}rgen \& Mathis (1994), we model the WIM of the
Galaxy as a combination of low-density \ion{H}{2} regions.  Our models
calculate the temperature and ionization structure of the model
\ion{H}{2} region from a distance of 0.3 pc from the exciting source to the 
point where the fraction of neutral hydrogen, $x(H^0)$, at the edge of
the nebula becomes larger than a critical value, $x_{edge}$.
Domg\"{o}rgen \& Mathis (1994) used models with $x_{edge}=0.10$ to
approximate fully-ionized material and those with $x_{edge}=0.95$ to
simulate ionized interfaces of neutral clouds or other regions with a
significant fraction of neutral hydrogen.  For continuity with their work,
we consider the same $x_{edge}$ values in this study. In
some situations, $x_{edge}$ values as high as 1.00 may be encountered as the 
sight line passes into a fully neutral cloud, but the 
model results for $x_{edge}=0.95$ closely approximate those for 
$x_{edge}=1.00$.  The emission
line intensities of the gas in these two cases are 
indistinguishable, and the ionization fractions vary by less than $\sim10$\% 
for most ions.  In this study we also consider a hybrid gas mixture described 
by a linear combination of equal amounts of $x_{edge}=0.10$ gas and 
$x_{edge}=0.95$ gas.  

The characteristics of a low-density photoionized nebula are uniquely
determined by three input parameters, assuming a given set of atomic
data.  These parameters are the set of gas-phase abundances adopted
for elements heavier than hydrogen, the shape of the ionizing
spectrum, and the ionization parameter, $q$, which is a measure of the
ratio of ionizing photons to particles in the nebula (see
Domg\"{o}rgen \& Mathis 1994 and Howk \& Savage 1999).  The inclusion of 
dust grain heating and cooling can also affect the model results, but
these effects are small in the models considered (see \S3.3).

We consider three sets of abundances in our models: solar system,
B-star, and warm neutral interstellar medium values, all of which are
summarized in Table~2.  The abundance of He is assumed to be one-tenth
the abundance of H in all three cases.  The solar system abundances
are meteoritic values from Anders \& Grevesse (1989) except for C, N,
and O, which are photospheric values from Grevesse \& Noels (1993).
The B-star abundances, which may represent a better ``cosmic''
reference system than solar system abundances (see, e.g., Savage \&
Sembach 1996 and Mathis 1996), are averages of the Kilian-Montenbruck,
Gehren, \& Nissen (1994) determinations.  We assume that the B-star 
abundances of elements lacking observed values are equal to the solar
values scaled downwards by 0.2 dex.  
We also consider a WNM-like abundance 
pattern to account for the
possible effects of elemental incorporation into dust grains in the
ionized interstellar medium.  The elemental depletion pattern inferred for
the WIM is comparable to that observed in the WNM (Howk \& Savage
1999; see also Lagache et al.  1999).  We adopt WNM gas-phase
abundances equal to those observed in the well-studied warm diffuse
clouds toward the low-halo star $\mu$~Columbae (Howk, Savage, \&
Fabian 1999).  The observed abundances in these clouds exhibit the
imprint of elemental incorporation into dust grains.  We adopt C, N,
and O abundances for the WNM from Sofia et al. (1997), Meyer et
al. (1997), and Meyer et al. (1998), respectively.  

Our models are unable to match the
[\ion{S}{2}]/[\ion{N}{2}] ratios observed by Haffner et al. (1999) if
we adopt a solar abundance for sulfur.  The predicted values of 
[\ion{S}{2}]/[\ion{N}{2}] are too high by roughly a factor of 1.5.  We have 
explored possible explanations for high [\ion{S}{2}]/[\ion{N}{2}] ratios
within the context of our model, but find that the most reasonable explanation
is a sub-solar reference abundance.  (Haffner et al. (1999) reached a similar
conclusion.)  The $\mu$~Columbae clouds exhibit nearly solar 
gas-phase abundances of the non-depleted elements S, Zn, and P, but this 
sight line may be unusual in this respect (Howk et al. 1999).  Therefore, 
for the elements S, Zn, P, Ar, and Na we adopt WNM reference abundances equal 
to the solar system values scaled downwards by 0.2 dex, in agreement with our 
current understanding of the abundances in the local ISM (Mathis 1996; Snow 
\& Witt 1996; Meyer et al. 1998).  This is essentially equivalent to adopting
a B-star reference abundance with minimal incorporation of these elements into
interstellar dust grains.

All of the \ion{H}{2} region models calculated with the adopted WNM
abundances include a standard mixture of interstellar graphite and
silicate grains (see Ferland 1996 and Baldwin et al. 1991).
Dust grains in the model nebulae affect the thermal
structure of the medium through photoelectric heating.  The grain
opacity also alters the shape of the ionizing spectrum slightly.  The
incorporation of potential coolants into grains is an essential component 
of the model and is required to match the observed emission-line strengths.

We use ATLAS line-blanketed LTE stellar atmosphere models (Kurucz
1991) as input spectra to the CLOUDY ionization code.  The input
spectral shape is varied by changing the effective temperature of the
stellar atmosphere of the central ionizing source.  We consider model
\ion{H}{2} regions ionized by central stars with stellar effective
temperatures in the range $31,000 \leq T_{eff} \leq 41,000$\,K.

The third input to our models, the ionization parameter, 
is a dimensionless measure of the relative
particle and ionizing photon densities.  Throughout this work we write
the ionization parameter, $q$, in the form

\begin{equation}
q \equiv n_{\rm H} f^2 L_{50},
\label{eqn:ionparam}
\end{equation}

\noindent
where $L_{50}$ is the stellar ionizing luminosity in units of
$10^{50}$ photons s$^{-1}$ and $f$ is the volume filling factor of the
ionized gas (Domg\"{o}rgen \& Mathis 1994). This definition of the
ionization parameter, $q$, is related to the ``volume averaged''
ionization parameter, $U = \frac{1}{3} \langle n_\gamma/n_e \rangle$,
through

\begin{equation}
q = (10^{-50}) \left( \frac{36 \pi c^3}{\alpha^2_B} \right) U^3 =
(1.33\times10^{-50}) \left( \frac{\pi c^3}{\alpha^2_B} \right) 
 \langle n_\gamma/n_e \rangle^3,
\end{equation}

\noindent
where $\alpha_B$ is the recombination coefficient of H to levels
n$\ge$2, and $n_\gamma$ is the number density of ionizing photons.

Bright high-density \ion{H}{2} regions typically have values of
$\log (q) \gtrsim -1.0$.  We calculate models over the range
$-4.0 \lesssim \log (q) \lesssim -3.0$, which is more appropriate for
the low-density WIM (Domg\"{o}rgen \& Mathis 1994).  For a nebula
surrounding a normal O9~V star with $T_{eff} = 35,000$ K, total
luminosity $\log L_* / L_\odot = 5.0$, and ionizing photon flux $\log
L_{50} \approx -1.6$ (Vacca, Garmany, \& Shull 1996), a value 
$\log (q) = -4.0$ corresponds to $n_e
f^2 \approx 4 \times 10^{-3}$ cm$^{-3}$.  Estimates for the density
of the WIM typically yield $f n_e \sim 0.08$ and $f \ga 0.20$ (Reynolds
1993).  Note that the ionization parameter rather than the density
determines the properties of the model nebulae; for a given value of
$q$ in Eq. (2), higher values of $n_e$ can be achieved provided there
is a proportional change in $n_\gamma$.

We extract from our models the volume-averaged intensities of various
emission lines relative to H$\alpha$, as well as the ionization
fractions, $x(X^i)$, of elements important for absorption-line studies
of the low-density Galactic ISM\footnotemark.  We use volume-weighted
averages rather than radial averages since a random, extended sight line
through the WIM is likely to sample gas found within a variety of low-density
ionized regions.  The predicted emission-line ratios can be compared directly 
with observations of the Galactic WIM (\S2), while the ionization fractions 
are useful for analyzing absorption-line data.

\footnotetext{All emission-line ratios in this paper are intensity ratios,
where the intensities are measured in energy units, not photon units.}
 
Several caveats regarding these models are worth mentioning.  First,
the atomic data incorporated in CLOUDY, while extensive, is
incomplete.  For example, the low-temperature dielectronic
recombination coefficients for most elements in the third and fourth row
of the periodic table are
estimated (Ferland 1996; Ferland et al. 1998). For some elements this
recombination pathway can be a significant contributor to the
ionization balance at nebular temperatures.  Although the atomic data
are somewhat incomplete for many of the elements considered, we
believe it is important to include the current best values of the
ionization fractions of all the elements that might be of interest.
Second, some of the emission-line predictions used to match the WIM
observations are highly sensitive to the predicted temperature of the
medium (Osterbrock 1989).  We discuss this further in \S3.3.  CLOUDY
calculates the thermal equilibrium structure of a model nebula, but
there may be additional heating and cooling processes in the WIM that
are not taken into account in the models. We assume that the WIM is
photoionized purely by stellar continuum radiation (as well as diffuse
radiation from the nebular gas itself).  Other processes may play a secondary
role in ionizing the WIM, such as magnetic reconnection, dissipation
of turbulence, or the mixing of cool material with hot, collisionally ionized 
gas (Raymond 1992; Slavin, Shull, \& Begelman 1993; Minter \& Balser 1997).

\subsection{The Standard Model}

The parameter space explored by our models spans a wide range of
physical properties in the ionized gas.  Our preferred model for the
WIM, hereafter referred to as the ``standard model,'' assumes WNM-like
abundances with interstellar grains included.  The ionizing spectrum
is that of a 35,000 K star, consistent with the upper limits to the
\ion{He}{1} $\lambda$5876 line emission set by observations (Reynolds
\& Tufte 1995).  We assume a very dilute radiation field 
characterized by $\log(q) = -4.0$.

Table~3 contains the emission-line intensities relative to H$\alpha$ predicted
by our CLOUDY models.  We give values for the $x_{edge} = 0.10$,
$x_{edge} = 0.95$, and standard composite models.  The composite model provides
a reasonable description of the observed emission-line ratios, with
the possible exceptions of [\ion{O}{1}] $\lambda6300$/H$\alpha$ and
[\ion{O}{3}]~$\lambda5007$/H$\alpha$.  The standard model produces
less [\ion{O}{3}] emission relative to H$\alpha$ than is suggested by
the observations.  This is a common feature of our models.  
The predicted [\ion{O}{1}]/H$\alpha$ ratio may also be slightly low
compared to the observed values (see Table~1).  While most directions
have observable values of [\ion{O}{1}]/H$\alpha \sim 0.02$, our composite 
models typically predict $\la 0.015$.  The strength of the 
[\ion{O}{1}] $\lambda6300$ line, however, is strongly
dependent on the temperature of the medium (Reynolds et al. 1998a).

We are not overly concerned about the possible under-production of
[\ion{O}{3}] $\lambda5007$ relative to H$\alpha$ in our models
compared to the observed upper limit of $\lesssim 0.1$.  This
observational limit was derived by noting that the current observed
values (Table~1) are at best 2--3$\sigma$ detections, so that the
actual value may be considerably lower than this limit.  Enhanced
[\ion{O}{3}] emission can be produced by alternate mechanisms (e.g.,
shocks) as well as by photoionization, so we expect that the observed
values should indeed provide a conservative upper bound for the model
predictions.

Previous investigations seeking to model the WIM emission-line
properties (e.g., Domg\"{o}rgen \& Mathis 1994) predicted higher
[\ion{O}{1}] $\lambda6300$ and [\ion{O}{3}] $\lambda5007$ emission
than our standard model.  To disentangle the differences between these
earlier works and our models, we have calculated CLOUDY models with
the abundances, ionizing spectrum, and ionization parameter adopted by
Domg\"{o}rgen \& Mathis.  We compare their predicted emission-line
ratios with those from our CLOUDY models in Table~4 and find good
agreement in most cases.  The slight remaining discrepancies,
particularly those for [\ion{O}{3}], are probably due to differences
in the atomic data used in the models (Ferland et al. 1998).  The
elements included in the models compared in Table~4 were limited to H,
He, N, O, Ne, and S.  As a result, the cooling is less efficient than
in our standard model, which incorporates elements up through Zn, and
the resulting nebular temperatures are higher in the
Domg\"{o}rgen \& Mathis models ($T_{neb}\sim8000$\,K versus
$T_{neb}\sim6700$\,K in our standard model).  An extensive comparison
of CLOUDY to other ionization codes can be found in Ferland et
al. (1994).

We present volume-averaged ionization fractions for the standard model
in Table~5. Values of $x(X^i)$ are listed for the first three
ionization stages of 20 elements.  Ionization potentials can be found
in Table~2, where we have underlined those ionization stages that have
at least one observable resonance line in the 912--3000\,\AA\ bandpass
suitable for high-resolution ultraviolet absorption-line work.  In the
standard composite model, hydrogen is almost fully ionized, $x(H^+) =
0.81$.  

\subsection{Sensitivity of Emission-Line Ratios to Model Parameters}

The emission-line ratios predicted by our models have varying
sensitivities to the input model parameters (abundance, ionizing
spectrum, ionization parameter).  In Figure~1 we illustrate the
dependences of the emission line ratios on the effective temperature
of the central ionizing source and the abundance pattern adopted for
models with $\log (q) = -4.0$.  The shaded regions indicate the ranges
of observed values as summarized in Table~3.  We have included
[\ion{O}{2}] $\lambda$3727 for completeness, though this line has not
yet been observed in the Galactic WIM.  Of the lines studied,
[\ion{S}{2}] $\lambda6716$ and [\ion{O}{1}] $\lambda6300$ are the most
sensitive to the adopted set of abundances.  \ion{He}{1} $\lambda5876$
and [\ion{O}{3}]~$\lambda5007$ are affected only slightly by the
chosen abundances, but vary by at least an order of magnitude over the
effective temperature range plotted.  [\ion{S}{2}] $\lambda6716$ and
[\ion{N}{2}] $\lambda6583$ are the least sensitive to the ionizing
spectrum, varying by less than a factor of 2 over the same range.

The upper limit to the \ion{He}{1}/H$\alpha$ ratio implies that the
ionizing spectrum of the WIM produces a ratio of He- to H-ionizing
photons less than that produced by an O star with $T_{eff}
\la$\,36,000 K (Reynolds \& Tufte 1995).  The observed 
[\ion{O}{1}]/H$\alpha$ constraint is barely satisfied over this range
of stellar effective temperatures, which indicates that there may be
some additional nebular heating required to increase the predicted
[\ion{O}{1}] emission.  Note, however, that [\ion{O}{1}] measurements
exist for a small number of WIM sight lines, and there appears to be a
significant variation in the [\ion{O}{1}]/H$\alpha$ ratio between these
directions.  The assumption that these emission line ratios are
representative of the WIM as a whole needs to be confirmed observationally.

The average nebular electron temperature in our standard model is
$T_{neb}=6700$\,K, which is determined by the input model parameters
and the detailed physics of heating and cooling in the nebula.  We
tabulate the effects of nebular temperature on the emission-line
ratios for the standard composite model in Table~6, where we have
forced $T_{neb}$ to be uniform throughout our model nebulae at
temperatures of 6000, 8000, and 10000\,K.  The recombination
line ratio \ion{He}{1}/H$\alpha$ is relatively unaffected by
nebular temperature, as expected, but the forbidden emission lines
vary greatly in strength as a function of $T_{neb}$. Changes of
$\sim500-1000$\,K in $T_{neb}$ can increase [\ion{O}{1}]/H$\alpha$ by
a factor of $\sim 2$.  In general, values of $T_{neb}$ as high as
8000~K are inconsistent with the observations in Table~3.  But, the observed
increase in [\ion{N}{2}]/H$\alpha$ and [\ion{S}{2}]/H$\alpha$
with height above the Galactic plane suggests temperatures as high as
$T_{neb}\sim10^4$ K may be appropriate at large $z$-distances
(Haffner et al. 1999).  Observations of [\ion{O}{2}] $\lambda3727$
could provide an excellent test for additional heating sources since
the [\ion{O}{2}]/H$\alpha$ ratio is expected to vary by a factor of
about~16 for $6000 \lesssim T_{neb} \lesssim 10000$\,K.

We note that grain heating in our models is approximately balanced by
grain cooling under the conditions considered.  The heating rate due to 
ejection of photoelectrons from grains is less than 10\% of the \ion{H}{1}
bound-free photoionization heating in our models. We refer the reader to 
Cox \& Reynolds (1992) and Baldwin et al. (1991) for additional discussions
of grain heating in the ionized ISM.

In general, the emission-line ratios do not depend strongly upon the
value of $q$ adopted for the standard model.  A value of $\log (q) =
-3.5$ versus $-4.0$ results in changes of less than 10\% for
[\ion{N}{2}]/H$\alpha$, [\ion{S}{2}]/H$\alpha$, [\ion{O}{1}]/H$\alpha$,
and \ion{He}{1}/H$\alpha$.  The [\ion{O}{3}]/H$\alpha$ ratio varies by
less than $\sim$40\%, with smaller changes occurring for values of
$T_{eff} < 38,000$\,K.

\subsection{Sensitivity of Ionization Fractions to Model Parameters}

The ionization fractions for the standard model (Table~5) have modest
dependencies on the input parameters used to model the emission line
observations.  We illustrate these dependencies in Figures~2 and~3,
where we plot the ionization fractions as functions of $T_{eff}$ and
$\log (q)$, respectively.  The standard $x_{edge}=0.10$,
$x_{edge}=0.95$, and composite models are shown with dashed, dotted,
and solid lines, respectively.  The ionization stages are
differentiated by the type of symbol used, with open circles for
neutral species, filled circles for singly ionized species, and
crossed circles for doubly ionized species.  For a few ions, the
ionization fractions depend strongly upon $T_{eff}$ (e.g., He\,{\sc
i-ii}, \ion{C}{3}), but in most other cases of interest here, the
variations with $T_{eff}$ are modest ($\Delta \log x(X^i) \le 0.2$
dex).  Similarly, changes in the adopted reference abundances do not
change the ionization fractions significantly.  The variations of 
$\log x(X^i)$ with $\log (q)$ shown in Figure~3 are typically less than 
0.2 dex for values
of $\log (q)$ between --4.0 and --3.0.

The ionization fractions do not depend significantly on the nebular
temperature since their values are governed in large part by
recombination processes.  Changes in $x(X^i)$ are less than
$\sim10$\% for $T_{neb}=6,000-10,000$\,K.  The lack of variation in the
strength of the permitted \ion{He}{1}~$\lambda5876$ recombination line
relative to H$\alpha$  with nebular temperature is
another manifestation of this insensitivity (see Table~6).

The information provided in Table~5 and Figures~2 and~3 should have a
broad range of applicability to studies of the WNM and WIM.  The ionization
fractions can be scaled easily to other preferred values of the input
model parameters.  As better emission-line data become available and
additional constraints are set on the properties of the WIM, the
standard model ionization fractions can also be refined using the
information presented here.
 
\section{Application to Absorption-Line Abundance Studies of the WNM and WIM}
\subsection{Methodology}

In this section we outline a simple strategy for estimating elemental
abundances in the warm neutral and ionized media of the Galaxy.  For
simplicity, any cold neutral gas along the sight line can be
considered as WNM material in the following discussion.  Many singly
ionized elements can be found in both the WNM and WIM (e.g.,
\ion{Al}{2}, \ion{Si}{2}, \ion{P}{2}, \ion{S}{2}, \ion{Fe}{2}), making it
necessary to account for the relative contributions of each type of
gas to the observed column densities.  Higher ionization stages are
preferentially found in the WIM, and these can be used to isolate the
elemental abundances in the WIM.

Assume that an ion $X^j$ is found only in the ionized gas and presumably 
has no appreciable abundance in the WNM (e.g., S$^{+2}$).  Then, we have

\begin{equation}
\begin{array}{l} N(X^i)_{WIM} = N(X^i)~~~~(i \ge j)
\\ N(X^i)_{WNM} = 0
\end{array} 
\end{equation}

\noindent
where $N(X^i)$ is the observed column density of $X^i$.  In general, the 
total amount of an ion is the sum of contributions from the WIM and WNM.

\begin{equation}
N(X^i) = N(X^i)_{WIM} + N(X^i)_{WNM}
\end{equation}

\noindent 
The column density of element $X$ in the WIM is given by

\begin{equation}
N(X)_{WIM} = N(X^j) / x(X^j)
\end{equation}

\noindent
where $x(X^j)$ is the volume-averaged ionization fraction found in Table~5.
The ionization fractions in Table~5 are appropriate for the average 
properties of the WIM.  The corresponding logarithmic abundance of the 
element relative to hydrogen in the WIM is 

\begin{equation}
A(X)_{WIM} = \log \left. \frac{N(X)}{N(H)}\right|_{WIM} + 12.00
\end{equation}

\noindent
It is difficult to estimate $N(H)_{WIM}$ directly since $N(H^+)$ is not 
directly observed, so a suitable proxy for $N(H)_{WIM}$ may need to be found.

In the WNM, the column density of element $X$ is equal to the observed 
value less the amount in the WIM, which is simply the sum of the individual 
contributions of $X^i$ to $N(X)$ in the WNM.

\begin{equation}
N(X)_{WNM} = \sum_{i=0}^{j-1} \left( N(X^i) - \frac{x(X^i)}{x(X^j)} N(X^j)
\right)
\end{equation}

\noindent
In Eq.(7), the second term in parentheses accounts for the amount of $X^i$
in the WIM.  
We use the following equations to estimate the amount of \ion{H}{1} in the WIM 
(Eq. 8) and WNM (Eq. 9):

\begin{equation}
F_H \approx 1 - \sum_{i=0}^{j-1} \left( \frac{N(X^i)_{WNM}}{N(X^i)} \right)
\end{equation}

\begin{equation}
N(H^0)_{WNM} = N(H^0) \left[ 1 - x(H^0) F_H \right]
\end{equation}

\noindent
In Eqs. (8) and (9), $F_H$ is the fraction of hydrogen along the sight line
associated with the WIM.  The ionization fraction of 
\ion{H}{1} in the WIM, $x(H^0)$, in our models is $\sim5-30\%$, so most
of the \ion{H}{1} is associated with the WNM.

Combining Eqs. (6) and (8), we derive the logarithmic abundance of X
relative to H in the WNM:

\begin{equation}
A(X)_{WNM} = \log \left. \frac{N(X)}{N(H^0)}\right|_{WNM} + 12.00
\end{equation}

\subsection{An Example: The HD\,93521 Sight Line}

To demonstrate our approach we apply the method outlined in
Eqs. (3)--(10) to estimate the gas-phase abundance of S in the warm
halo clouds towards HD~93521.  Spitzer \& Fitzpatrick (1993) have
presented high-resolution observations of this sight line obtained with
the Goddard High Resolution Spectrograph on-board the HST.  The
neutral and ionized species occur at very similar velocities along this sight
line, so the contributions of ionized gas to the derived elemental
abundances may be important.  Spitzer \& Fitzpatrick (1993)
measured the column densities of \ion{S}{2}, 
\ion{S}{3}, and \ion{H}{1} in the individual ``warm fast'' clouds along this 
sight line.  In the first five columns of Table~7 we give their
identifications, LSR velocities, and column densities of
\ion{S}{2}, \ion{S}{3}, and \ion{H}{1} for each of these clouds.  
These values are followed in column 6
by the column density of \ion{S}{2} in the WIM calculated using
Eq. (7) and the model predictions given in Table~5.  The ratio of
\ion{S}{2} in the WNM to total \ion{S}{2} in each cloud 
is listed in column 7.  Equations (8) and (9) were used to derive the
ratio of \ion{H}{1} in the WNM to the observed value.  We list this ratio
in column 8.  Column 9 contains
 the derived abundance of S in the WNM found from Eq. (10), and column 10
contains $A(S) - A_c(S)$, the logarithmic gas-phase abundance of sulfur 
relative to the WNM reference abundance given in Table~2.

Table~7 shows the results of these calculations using both the
standard composite model and the $x_{edge} = 0.10$ model results for
$x(X^i)$.  The cloud-to-cloud scatter in the values of $A(S)-A_c(S)$
is generally of the order of the measurement errors on
$N$(\ion{S}{3}).  Values of $A(S)-A_c(S)$ for components 1,2, and 4
for the standard composite model are consistent with a negligible
gas-phase depletion of S onto dust grains. The value for component 3
suggests either that there may be a small amount of S incorporated into dust
or that the ionization properties or intrinsic abundances of
cloud 3 vary enough to produce the observed deviation from the
reference abundance.  In the $x_{edge} = 0.10$ model, the --0.14 dex deviation
from the reference value is a 2$\sigma$ excursion.  The cloud 3 value
is 0.2 dex below the value of $A(S)$ derived by Spitzer \& Fitzpatrick (1993), 
who assumed that all of the \ion{S}{2} in these clouds is associated with 
the \ion{H}{1} (i.e., they did not apply any ionization corrections).

We find that a substantial ionization correction {\it is} necessary 
for \ion{S}{2} to explain the amount of \ion{S}{3} observed along the
sight line. We find that roughly 50\% of the \ion{S}{2} is in the WNM, with 
the other 50\% contained in the WIM (see Table~7).  It is interesting that 
Spitzer \& Fitzpatrick (1993) 
found an electron density for the HD\,93521 clouds comparable to that of the 
WIM.  The main difference in addressing ionization in these two studies is 
that they used the observed ratio N(\ion{S}{2})/N(\ion{S}{3}) and measured
electron density to derive the photoionization rate for \ion{S}{2} under 
conditions of photoionization equilibrium, whereas 
our results take into account the explicit ionizing spectrum required to match
the WIM properties.  We also predict varying ionization conditions between
clouds, whereas they assumed constant properties.  Spitzer \& Fitzpatrick 
(1993) argued that the free electrons and neutral gas along the HD\,93521 
sight line are well mixed.  Our calculation is consistent with this result, 
but does not require it.
Additional observations of other ionized gas species in the HD\,93521 clouds
would help to characterize their ionization conditions and the impact of 
ionization on the derived elemental abundances.

Analyses such as these have also been used successfully to study the
abundances in the WNM of the Galactic disk and halo and in
intermediate-velocity clouds (e.g., Sembach 1995; 
Howk et al. 1999). Eventually, similar studies of the WIM abundances
could be made (see Howk \& Savage 1999).  We note that especially good 
indicators of partially
ionized gas are observable at the far-ultraviolet wavelengths to be
observed by FUSE.  
Information about spectral lines that are useful probes of
photoionized gas can be found in Sembach (1999) and Howk \& Savage
(1999) - see Table~1 in both papers.

\section{Concluding Remarks}
The outlook for a better understanding of
the abundances and physical properties
of the WIM is bright.  Advances in
measuring the emission-line properties of the WIM will probably
occur at optical and near-infrared wavelengths in the near
future.  Emission-line studies of the WIM at wavelengths
$\lambda < 3000$\,\AA\ will require sensitive detectors with large
fields-of-view.  The predicted intensities of the strongest emission lines at
far-ultraviolet wavelengths ($\lambda < 1200$\,\AA) in our models are
currently below the sensitivity of FUSE. In the standard composite model 
[\ion{C}{2}] $\lambda1020$, \ion{C}{3}
$\lambda977$, \ion{N}{2} $\lambda1085$, and \ion{N}{3} $\lambda990$
all have intensities less than $10^{-2} I_{H\alpha}$.  At these strengths,
the far-ultraviolet WIM emission lines measured by FUSE in its
30\arcsec$\times$30\arcsec\ aperture should fall well below the
expected background levels of $\sim0.5$ counts sec$^{-1}$ cm$^{-2}$
and should pose no significant contamination problem for studies of
hot gas emission (e.g., \ion{O}{6}) from the highly ionized ISM.  
Absorption line studies of the WIM with FUSE should be feasible, with
excellent diagnostics of photoionized gas ranging in ionization energies 
from 1 to
4 Rydbergs (e.g., C\,{\sc ii-iii},
N\,{\sc i-iii}, Ar\,{\sc i-ii}, Fe\,{\sc ii-iii}, P\,{\sc ii-iv},
S\,{\sc iii-iv}).

The results of this work are applicable to warm ionized gas 
regions in other galaxies.  
A companion study of warm ionized gas and its effects on elemental abundance
determinations in damped Lyman-$\alpha$ absorbers can be found in Howk 
\& Sembach (1999).

\acknowledgments  
We thank Gary Ferland and his co-workers at the University of Kentucky
for all of their work on the CLOUDY code.  We also thank John Mathis for 
useful discussions about modeling the WIM and Ron Reynolds for helpful 
suggestions for improving the manuscript.
KRS and JCH acknowledge support from NASA Long Term Space Astrophysics grant 
NAG5-3485 and grant GO-07270.01-96A from the Space Telescope Science Institute,
which is operated by the Association of Universities for Research in
Astronomy, Inc., under NASA contract NAS5-26555.

\appendix

\section{Ionization Corrections for Low-Density H\,{\sc ii} Regions}

Although the main purpose of this work is to determine ionization
fractions for the WIM of the Galaxy, our CLOUDY
calculations also have applicability to studies of ionized gas in low-density
\ion{H}{2} regions.  \ion{H}{2} regions can be a significant source of
singly-ionized species, which are also the primary tracers of many elements
in the neutral ISM. Howk et
al. (1999) have discussed the effects of \ion{H}{2} 
region gas on the the study of the gas-phase abundances in the WNM along
the $\mu$ Columbae sight line.

The data given in Table~5 describe the volume-averaged properties of
our model nebulae.  For a sight line through a diffuse \ion{H}{2}
region toward the central ionizing source, it may be more appropriate to
use radially-averaged ionization fractions (see Howk \& Savage 1999).  In
Table~8 we list the radially-averaged ionization fractions for all 20
elements considered in this study.  We provide ionization fractions for
diffuse \ion{H}{2} regions characterized by $\log \, (q) = -3.0$ and
$-4.0$, with an ionizing spectrum from an ATLAS model atmosphere with
$T_{eff} = 35,000$ K.  We present model results for
both $x_{edge} = 0.10$ and 0.95.  Figures~4a and 4b show the
variations in these ionization fractions as a function of input
stellar effective temperature for a select number of elements for the
$\log \, (q) = -4.0$ models.  For the elements Al, Si, Mg, Cr, and Mn,
the variation in the ionization fractions of the singly and
doubly ionized stages is less than 0.05 dex over the range $31,000
\leq T_{eff} \leq 41,000$~K.  Figure~5 shows the variation in
ionization fraction with $\log \, (q)$, assuming an input spectrum
with $T_{eff} = 35,000$ K.

Howk \& Savage (1999) have tabulated radially-average ionization
fractions for a limited number of ions and discussed in detail how
the ionization fractions vary with assumed model parameters.  The
results presented below are more thorough in the sense that more
elements are covered, though over a smaller range of ionization
parameter.  These ionization fractions are appropriate for very
low-density \ion{H}{2} regions.  In particular, these results should
be useful for studying contamination from \ion{H}{2} regions surrounding 
early-type stars at large distances from the Galactic plane
or along short, low-density sight lines where an appreciable number of 
additional \ion{H}{2} regions has not been intercepted. 

The application of the ionization fractions given in Table~8 should be
approached in a method similar to that described in \S 4.  When
possible, multiple observations of adjacent ions should be used to
constrain the most appropriate model parameters.  The
radially-averaged values tend to favor higher-ionization gas than the
volume-averaged results, since the partially-ionized outer regions of
the nebula are given less weight in the averaging process.

\newpage
\begin{center}
{\bf Figure Captions}
\end{center}

\noindent
\figcaption{Predicted emission line ratios versus the assumed 
effective temperature of the underlying stellar ionizing source from the CLOUDY
model nebula calculations.  The three curves in each panel depict the
relationships for the different abundance patterns indicated in the
legend in the lower left panel.  For each set of reference abundances 
we show the composite model with an ionization parameter $\log (q) = -4.0$.
Shaded regions indicate the range of observed emission-line ratios
(see Table~1).  Note that [\ion{O}{2}] $\lambda3727$ emission has not
yet been observed in the WIM.}

\figcaption{Volume-averaged ionization fractions, $\log x(X^i)$, versus input 
stellar effective temperature of the ionizing central source in our nebular
models with WNM abundance patterns.  The curves shown are for the
$x_{edge} = 0.10$ (dashed lines), $x_{edge} = 0.95$ (dotted lines),
and composite (solid lines) cases.  The ionization stage (I-III) is
indicated by the type of symbol plotted: open circles (I),
filled circles (II), and crossed circles (III). \nl a)
He, C, N, and O; b) S, P, Ar, and Zn; c) Si, Mg, Mn, and Cr; d) Fe,
Ti, Al, and Ni.}

\figcaption{Volume-averaged ionization fractions, $\log x(X^i)$, versus 
ionization
parameter, $\log (q)$, for a model nebula with a WNM abundance pattern
and an ionizing source characterized by $T_{eff} = 35,000$ K.  The
curves shown are for the $x_{edge} = 0.10$ (dashed lines), $x_{edge} =
0.95$ (dotted lines), and composite (solid lines) cases.  The
ionization stage (I-III) is indicated by the type of symbol plotted:
open circles (I), filled circles (II), and crossed
circles (III).}

\figcaption{Radially-averaged ionization fractions, $\log x(X^i)$,
versus effective temperature assumed for the ionizing central star.
The curves shown are for the $x_{edge} = 0.10$ (dashed lines),
$x_{edge} = 0.95$ (dotted lines) models.  All models assume WNM
abundance patterns, but the ionization fractions are relatively
insensitive to the adopted abundances.  The ionization stage (I-III)
is indicated by the type of symbol plotted: open circles (I),
filled circles (II), and crossed circles (III). \nl a)
C, N, O, and Zn; b) S, P, Fe, and Al.}

\figcaption{Radially-averaged ionization fractions, $\log x(X^i)$,
versus ionization parameter, $\log \, (q)$. The curves shown are for
the $x_{edge} = 0.10$ (dashed lines), $x_{edge} = 0.95$ (dotted lines)
models.  These models assume an ionizing central star with $T_{eff} =
35,000$ K.  All models assume WNM abundance patterns, but the
ionization fractions are relatively insensitive to the adopted
abundances.  The ionization stage (I-III) is indicated by the type of
symbol plotted: open circles (I), filled circles (II),
and crossed circles (III).}

\end{document}


\newcommand{\degr}{$^{\circ}$}
\newcommand{\kms}{\,km\,s$^{-1}$}
\newcommand{\updag}{$^\dagger$}     
 
\begin{deluxetable}{ccccccc}
\tablecaption{Emmission Line Measurements of the Warm Ionized Medium}
\tablecolumns{7}
\tablehead{
Direction 
& $\frac{[S\,\sc{ii}]~\lambda6716}{H\alpha}$
& $\frac{[N\,\sc{ii}]~\lambda6583}{H\alpha}$
& $\frac{[O\,\sc{i}]~\lambda6300}{H\alpha}$
& $\frac{[O\,\sc{iii}]~\lambda5007}{H\alpha}$
& $\frac{He\,\sc{i}~\lambda5876}{H\alpha}$
& Reference\tablenotemark{a} \nl
($l$, $b$)}
\startdata
\phn96\fdg0, 0\fdg0   &  0.29$\pm$0.05   &   0.28$\pm$0.06   &  \nodata         &  0.058$\pm$0.025  &  \nodata & 1,2 \nl
\nl
114\fdg0, 0\fdg0       &  0.24$\pm$0.04   &   \nodata        &  0.020$\pm$0.003  &  \nodata         &  $<$0.02 & 2,3,4 \nl
\nl
130\fdg0, 0\fdg0       &  0.27$\pm$0.06   &   \nodata        &  0.028$\pm$0.009  &  \nodata         &  \nodata & 2,3 \nl
\nl
$\left. \begin{array}{c} 184\fdg4, -0\fdg6 \\ 186\fdg0, 0\fdg0 \end{array} \right\} $
	                &  0.23$\pm$0.04   &  0.21$\pm$0.02    &  \nodata          &  $<$0.06         &  \nodata & 2,5 \nl
\nl
194\fdg0, 0\fdg0       &  0.38$\pm$0.05   &  0.44$\pm$0.05    &  \nodata          &  0.068$\pm$0.025 &  $<$0.02 & 1,2,4 \nl
\nl
$\left. \begin{array}{c} 97\fdg3, -29\fdg7 \\ 97\fdg7, -26\fdg7 \end{array} \right\} $ 
   	                &  0.39$\pm$0.11   &  \nodata         &  \nodata          &  $<$0.20         &  \nodata & 1,2 \nl
\nl
144\fdg0, --21\fdg0    &  0.44            &  0.70             &  $<0.12$           &  $<$0.57         &  \nodata & 6 \nl
{ \small (weak shock)} \nl
\nl
123 to 164, --6 to --35 &  0.2--0.4\tablenotemark{b}        &  0.4--0.6\tablenotemark{b}         
&  $\left\{\begin{array}{c} <0.012\tablenotemark{c} \\ 0.044\pm0.011\tablenotemark{c} \end{array} \right\}$          
&  \nodata        &  \nodata & 3,7\nl
{\small (Perseus Arm)} \nl

\enddata
\tablenotetext{a}{References: (1) Reynolds 1985b, (2) Reynolds 1985a, (3) Reynolds et al. 1998a, 
(4) Reynolds \& Tufte 1995, (5) Reynolds, Roessler, \& Scherb 1977; (6) Ogden \& Reynolds (1985), 
(7) Haffner, Reynolds, \& Tufte 1999.}
\tablenotetext{b}{The ranges of values quoted are appropriate for Perseus Arm gas within
$\la 1$ kpc of the Galactic plane having an H$\alpha$ intensity of $\sim 2-3$~R.  
There is a slightly larger scatter to higher 
values in the Local Arm gas in the same directions.
Both the [\ion{S}{2}]/H$\alpha$ and [\ion{N}{2}]/H$\alpha$ ratios are anti-correlated with I$_{H\alpha}$.
The upper envelope of observed ratios extends to values several times higher than those listed.}
\tablenotetext{c}{These values are for two velocities (--60 and --31 km~s$^{-1}$) at the position 
$l = 130$\degr, $b = -7.5$\degr.}

\end{deluxetable}

\begin{deluxetable}{lccclll}
\tablewidth{400pt}
\tablecolumns{7}
\tablecaption{Ionization Potentials and Reference Abundances}
\tablehead{Element   &  \multicolumn{3}{c}{Ionization Potentials (eV)\tablenotemark{a}} 
& \multicolumn{3}{c}{$A(X)$\tablenotemark{b}} \nl
\cline{2-4}
\cline{5-7}
& I & II & III & Solar  &   B-star &  WNM}
\startdata
H    & 13.60 & \nodata & \nodata & 12.00 & 12.00 & 12.00 \nl
He   & 24.59 & 54.42 & \nodata & 11.00  &    11.00        &  11.00 \nl
C    & 11.26 & \underline{24.38} & \underline{47.89}   & \phn8.55  &  \phn8.27         &  \phn8.15   \nl
N    & \underline{14.53} & \underline{29.60} & \underline{47.45}   & \phn7.97  &    \phn7.80         &  \phn7.88 \nl
O    & \underline{13.62} & 35.12 & 54.93   &  \phn8.87  &    \phn8.56           &  \phn8.50 \nl
Ne   & 21.56 & 40.96 & 63.45   &  \phn8.07  &    \phn8.20            &  \phn8.07 \nl
Na   & \phn5.14 & 47.29 & 71.64   &  \phn6.31  &    \phn6.11$^\dagger$  &  \phn6.11 \nl
Mg   & \phn\underline{7.65} & \underline{15.04} & 80.14   &  \phn7.58  &    \phn7.38            &  \phn7.16 \nl
Al   & \phn5.99 & \underline{18.83} & \underline{28.45}   &  \phn6.48  &    \phn6.19            &  \phn5.18 \nl
Si   & \phn8.15 & \underline{16.34} & \underline{33.49}  &  \phn7.55  &    \phn7.20            &  \phn7.18 \nl
P    & 10.49 & \underline{19.72} & \underline{30.18}   &  \phn5.57  &    \phn5.37$^\dagger$  &  \phn5.37 \nl
S    & 10.36 & \underline{23.33} & \underline{34.83}   &  \phn7.27  &    \phn6.97            &  \phn7.07 \nl
Ar   & \underline{15.76} & \underline{27.63} & 40.74   &  \phn6.52  &    \phn6.32$^\dagger$  &  \phn6.32 \nl
Ca   & \phn6.11 & 11.87 & 50.91   &  \phn6.36  &    \phn6.16$^\dagger$  &  \phn5.06 \nl
Ti   & \phn6.82 & 13.58 & 27.49   &  \phn4.93  &    \phn4.73$^\dagger$ & \phn3.86 \nl
Cr   & \phn6.77 & \underline{16.50} & \underline{30.96}   &  \phn5.68  &    \phn5.48$^\dagger$  &  \phn4.55 \nl
Mn   & \phn7.43 & \underline{15.64} & 33.67   &  \phn5.53  &    \phn5.33$^\dagger$  &  \phn4.56 \nl
Fe   & \phn7.87 & \underline{16.18} & \underline{30.65}   &  \phn7.51  &    \phn7.40            &  \phn6.21 \nl
Ni   & \phn7.64 & \underline{18.17} & 35.17   &  \phn6.25  &    \phn6.05$^\dagger$  &  \phn4.95 \nl
Zn   & \phn9.39 & \underline{17.96} & 39.72   &  \phn4.65  &    \phn4.45$^\dagger$  &  \phn4.45 \nl

\enddata
\tablenotetext{a}{Ionization potentials are from Moore (1970).  Underlined values
indicate that the ionization stage has at least one ground-state 
resonance line in the FUSE/HST ultraviolet 
bandpass (912--3000\,\AA) that could be observable in the WIM.}
\tablenotetext{b}{Reference abundances are logarithmic values on a scale
where $A(X) = \log (X/H) + 12.00$.
Solar abundances are meteoritic
values from Anders \& Grevesse (1989), except for C, N, and O, which are
photospheric values from Grevesse \& Noels (1993).  B-star abundances 
are from Kilian-Montenbruck et al. (1994), except for those cases marked with 
a $\dagger$, which are equal to the solar value minus 0.2 dex.  The WNM
(warm neutral medium) abundances are equal to the solar abundances modified by
dust as observed in the warm diffuse clouds toward $\mu$~Columbae (Howk et al.
1999). WNM values for C, N, and O are from Sofia et al. (1997), Meyer et al.
(1997), and Meyer et al. (1998), respectively.}
%
%
\end{deluxetable}

\clearpage
\begin{deluxetable}{lcccc}
\tablecolumns{5}
\tablewidth{0pt}
\tablecaption{Standard Model WIM Emission Lines}
\tablehead{Line & \multicolumn{3}{c}{Volume-Averaged Intensity Ratio ($\frac{I_\lambda}{I_{H\alpha}}$)} 
& Observed\tablenotemark{a} \nl
\cline{2-4}
& $x_{edge} = 0.10$ & $x_{edge} = 0.95$ & Composite\tablenotemark{b} }
\startdata
[\ion{S}{2}] $\lambda6716$ 	& 0.258 & 0.325 & 0.292 & 0.2--0.4\nl
\nl
[\ion{N}{2}] $\lambda6583$	& 0.439 & 0.462 & 0.451 & 0.3--0.6\nl
\nl
[\ion{O}{1}] $\lambda6300$	& 0.005 & 0.021 & 0.013 & 0.01--0.03\nl
\nl
[\ion{O}{3}] $\lambda5007$	& 0.002 & 0.002 & 0.002 & $\lesssim$0.1\nl
\nl
\ion{He}{1}  $\lambda5876$	& 0.017 & 0.013 & 0.015 & $<$0.02\nl
\enddata
\tablenotetext{a}{This summary of observed intensity ratios is based on the 
data listed in Table~1.}
\tablenotetext{b}{The composite model is comprised of equal amounts of 
$x_{edge} = 0.10$ and $x_{edge} = 0.95$ material.}
\end{deluxetable}

\clearpage 
\begin{deluxetable}{lcc}
\tablewidth{200pt}
\tablecaption{Emission Line Comparison}
\tablehead{Line   &  \multicolumn{2}{c}{Intensity Ratio\tablenotemark{a}~~($\frac{I_\lambda}{I_{H\alpha}}$)}\nl
\cline{2-3}
& CLOUDY  &  DM94}
\startdata
[\ion{S}{2}] $\lambda6716$ &  0.77   &  0.66	\nl
[\ion{N}{2}] $\lambda6583$ &  0.37   &  0.37	\nl
[\ion{O}{1}] $\lambda6300$ &  0.051  &  0.055 	\nl
[\ion{O}{2}] $\lambda3727$ &  1.17   &  1.18	\nl
[\ion{O}{3}] $\lambda5007$ &  0.009  &  0.038	\nl
\ion{He}{1} $\lambda5876$  &  0.025  &  0.029	\nl
\nl
$T_{neb}$(K)\tablenotemark{b}   & 8040 &  8200 \nl
\enddata
\tablenotetext{a}{These volume-averaged intensity ratios
are predictions from the CLOUDY and Domg\"{o}rgen \& Mathis (1994)
ionization codes for a 
model nebula with $x_{edge} = 0.95$, $T_{eff}$ = 38,000 K, and
$\log (q) = -4.0$.  The calculations included
H, He, N, O, Ne, and S scaled to the Orion abundances from 
Peimbert et al. (1992).}
\tablenotetext{b}{Volume averaged nebular temperature 
defined by $T_{neb} = \int{T n_e dV} / \int{n_e dV}$.}
\end{deluxetable}

\clearpage

\begin{deluxetable}{lccccccccccc}
\tablewidth{0pt}
\tablecaption{Ionization Fractions for the Standard Model\tablenotemark{a,b}}
\tablehead{Element   &  \multicolumn{3}{c}{$x_{edge} = 0.10$} &&   \multicolumn{3}{c}{$x_{edge} = 0.95$} &&  
\multicolumn{3}{c}{Composite\tablenotemark{c}} \nl
\cline{2-4}
\cline{6-8}
\cline{10-12}
  & I & II  & III   && I & II  & III  && I & II  & III}
\startdata
H     &  --1.30  &  --0.02  &  \nodata & &    --0.49  &    --0.17  &   \nodata & &    --0.73  &    --0.09  &   \nodata  \nl
He    &  --0.17  &  --0.50  &  \nodata & &    --0.07  &    --0.83  &   \nodata & &    --0.12  &    --0.63  &   \nodata  \nl
C     &  --2.24  &  --0.02  &  --1.31  & &    --2.16  &    --0.01  &    --1.63 & &    --2.20  &    --0.02  &    --1.44  \nl  
N     &  --1.33  &  --0.03  &  --1.72  & &    --0.52  &    --0.16  &    --2.05 & &    --0.76  &    --0.09  &    --1.85  \nl   
O     &  --1.26  &  --0.03  &  --2.74  & &    --0.48  &    --0.18  &    --3.07 & &    --0.71  &    --0.09  &    --2.87  \nl   
Ne    &  --0.97  &  --0.05  &  --3.88  & &    --0.54  &    --0.15  &    --4.21 & &    --0.70  &    --0.10  &    --4.02  \nl   
Na    &  --1.93  &  --0.01  &  --5.53  & &    --1.75  &    --0.01  &    --5.82 & &    --1.83  &    --0.01  &    --5.65  \nl   
Mg    &  --2.30  &  --0.15  &  --0.54  & &    --2.22  &    --0.07  &    --0.82 & &    --2.26  &    --0.11  &    --0.66  \nl   
Al    &  --2.81  &  --0.03  &  --1.16  & &    --2.83  &    --0.02  &    --1.44 & &    --2.82  &    --0.03  &    --1.28  \nl   
Si    &  --3.63  &  --0.06  &  --0.92  & &    --3.51  &    --0.03  &    --1.23 & &    --3.57  &    --0.04  &    --1.05  \nl   
P     &  --2.39  &  --0.09  &  --0.72  & &    --2.36  &    --0.05  &    --1.01 & &    --2.37  &    --0.07  &    --0.84  \nl   
S     &  --2.83  &  --0.11  &  --0.65  & &    --2.76  &    --0.06  &    --0.90 & &    --2.80  &    --0.09  &    --0.75  \nl   
Ar    &  --1.86  &  --0.12  &  --0.63  & &    --0.94  &    --0.11  &    --0.96 & &    --1.19  &    --0.12  &    --0.77  \nl   
Ca    &  --2.33  &  --0.71  &  --0.10  & &    --2.31  &    --0.65  &    --0.11 & &    --2.32  &    --0.68  &    --0.10  \nl   
Ti    &  --3.19  &  --0.34  &  --0.28  & &    --2.87  &    --0.15  &    --0.54 & &    --3.00  &    --0.23  &    --0.39  \nl   
Cr    &  --3.42  &  --0.41  &  --0.21  & &    --3.12  &    --0.17  &    --0.48 & &    --3.24  &    --0.28  &    --0.33  \nl   
Mn    &  --1.68  &  --0.31  &  --0.31  & &    --1.39  &    --0.16  &    --0.58 & &    --1.51  &    --0.23  &    --0.43  \nl   
Fe    &  --3.41  &  --0.47  &  --0.19  & &    --3.02  &    --0.19  &    --0.45 & &    --3.17  &    --0.31  &    --0.30  \nl  
Ni    &  --1.98  &  --0.11  &  --0.68  & &    --1.89  &    --0.05  &    --1.00 & &    --1.93  &    --0.08  &    --0.81  \nl  
Zn    &  --2.15  &  --0.04  &  --1.08  & &    --1.96  &    --0.02  &    --1.40 & &    --2.05  &    --0.03  &    --1.21  \nl  
\enddata
\tablenotetext{a}{The ionization fraction of an ion $X^i$ is given by 
	$x(X^i) = N(X^i)/N(X_{\rm Total})$.  We present volume-averaged values of
	$\log x(X^i)$ for the first three ionization states of each element.}
\tablenotetext{b}{The standard model assumes WNM abundances (see Table 1), 
	log\,$(q)$ = --4.0, and a stellar input spectrum characterized by $T_{eff}$ = 35,000 K and 
	log\,(g) = 4.0.}
\tablenotetext{c}{The composite model is described by 50\%($x_{edge} = 0.10$) + 50\%($x_{edge} = 0.95$).}
\end{deluxetable}

\clearpage
\begin{deluxetable}{lccc}
\tablecolumns{4}
\tablewidth{280pt}
\tablecaption{Effects of Nebular Temperature on Emission Lines}
\tablehead{Line & \multicolumn{3}{c}{Intensity Ratio\tablenotemark{a}~~($\frac{I_\lambda}{I_{H\alpha}}$)} \nl
\cline{2-4}
& 6,000\,K & 8,000\,K & 10,000\,K}
\startdata
[\ion{S}{2}] $\lambda6716$ &  0.200  &  0.539  &  0.990   \nl  
\nl 
[\ion{N}{2}] $\lambda6583$ &  0.291  &  0.820  &  1.571   \nl  
\nl  
[\ion{O}{1}] $\lambda6300$ &  0.009  &  0.032  &  0.075   \nl  
\nl 
[\ion{O}{3}] $\lambda5007$ &  0.001  &  0.002  &  0.005   \nl 
\nl  
\ion{He}{1}  $\lambda5876$ &  0.016  &  0.015  &  0.014   \nl   
\nl
[\ion{O}{2}] $\lambda3727$ &  0.186  &  1.050  &  3.031   \nl   
\nl
[\ion{S}{3}] $\lambda9532$ &  0.085  &  0.173  &  0.271   \nl   
\enddata
\tablenotetext{a}{These volume-averaged intensity ratios are
for the standard composite model, which is comprised of equal amounts of 
$x_{edge} = 0.10$ and $x_{edge} = 0.95$ material.}
\end{deluxetable}

\clearpage
\begin{deluxetable}{cccccccccc}
\tablecolumns{10}
\tablewidth{550pt}
\tablecaption{Standard Model Results for the HD\,93521 Sight Line\tablenotemark{a}}
\tablehead{
Comp. & $<V_{LSR}>$ & $\log N$(\ion{S}{2}) & $\log N$(\ion{S}{3}) & $\log N$(\ion{H}{1}) 
& $\log N$(\ion{S}{2})$_{wim}$ & $\frac{N(S\,{\sc ii})_{wnm}}{N(S\,{\sc ii})}$ 
& $\frac{N(H\,{\sc i})_{wnm}}{N(H\,{\sc i})}$ & A(S) & A(S)--A$_c$(S) \nl
& (km~s$^{-1}$)}
\startdata
\cutinhead{Composite Model}
1 & $-$66.3 & 14.09$\pm$0.03 & 13.05$\pm$0.21 & 18.85 & 13.71 & 0.58 & 0.92 & 7.04 & $-$0.03 \nl
2 & $-$57.8 & 14.50$\pm$0.02 & 13.11$\pm$0.22 & 19.26 & 13.77 & 0.81 & 0.97 & 7.17 & $+$0.10 \nl
3 & $-$51.2 & 14.53$\pm$0.03 & 13.71$\pm$0.07 & 19.29 & 14.37 & 0.31 & 0.87 & 6.79 & $-$0.28 \nl
4 & $-$38.8 & 14.02$\pm$0.07 & 13.14$\pm$0.31 & 18.78 & 13.80 & 0.40 & 0.89 & 6.89 & $-$0.18 \nl
Sum & \nodata & 14.95$\pm$0.02 & 13.95$\pm$0.08 & 19.71 & 14.61 & 0.54 & 0.91 & 7.01 & $-$0.06 \nl
\cutinhead{$x_{edge} = 0.10$}
1 & $-$66.3 & 14.09$\pm$0.03 & 13.05$\pm$0.21 & 18.85 & 13.59 & 0.68 & 0.94 & 7.08 & $+$0.01 \nl
2 & $-$57.8 & 14.50$\pm$0.02 & 13.11$\pm$0.22 & 19.26 & 13.65 & 0.86 & 0.97 & 7.18 & $+$0.11 \nl
3 & $-$51.2 & 14.53$\pm$0.03 & 13.71$\pm$0.07 & 19.29 & 14.25 & 0.48 & 0.90 & 6.93 & $-$0.14 \nl
4 & $-$38.8 & 14.02$\pm$0.07 & 13.14$\pm$0.31 & 18.78 & 13.68 & 0.54 & 0.91 & 6.98 & $-$0.09 \nl
Sum & \nodata & 14.95$\pm$0.02 & 13.95$\pm$0.08 & 19.71 & 14.49 & 0.65 & 0.94 & 7.06 & $-$0.01 \nl
\enddata
\tablenotetext{a}{Observed quantities ($<V_{LSR}>$, $N$(\ion{S}{2}), $N$(\ion{S}{3}), $N$(\ion{H}{1}))
are measurements for the ``warm, fast'' halo clouds observed toward HD\,93521 by Spitzer \& Fitzpatrick 
(1993).  The remaining derived quantities are defined in the text.}
\end{deluxetable}

\clearpage 
\begin{deluxetable}{lccccccccccccccc}
\tablewidth{0pt}
\tablecaption{Ionization Fractions for Low-Density H\,{\sc ii} Regions\tablenotemark{a}}
\tablehead{
          &  \multicolumn{7}{c}{$\log \, (q) = -3.0$}  & & \multicolumn{7}{c}{$\log \, (q) = -4.0$}  \nl
\cline{2-8} 
\cline{10-16}
Element   &  \multicolumn{3}{c}{$x_{edge} = 0.10$} &&   \multicolumn{3}{c}{$x_{edge} = 0.95$} &&  
   	     \multicolumn{3}{c}{$x_{edge} = 0.10$} &&   \multicolumn{3}{c}{$x_{edge} = 0.95$} \nl
\cline{2-4}
\cline{6-8}
\cline{10-12}
\cline{14-16}
      & I       & II       & III     &&   I      & II      & III      &&    I     &  II     &  III     &&   I       & II       & III}
\startdata
H     & --1.81  &  --0.01  & \nodata & & --1.08  & --0.04  &  \nodata & & --1.60  & --0.01  &  \nodata & &  --0.87  &  --0.06  &  \nodata \nl
He    & --0.44  &  --0.20  & --8.06  & & --0.34  & --0.26  &  --8.13  & & --0.42  & --0.21  &  --7.52  & &  --0.28  &  --0.32  &  --7.63   \nl
C     & --2.81  &  --0.17  & --0.50  & & --2.71  & --0.14  &  --0.57  & & --2.52  & --0.12  &  --0.61  & &  --2.37  &  --0.10  &  --0.72   \nl  
N     & --1.87  &  --0.14  & --0.59  & & --1.13  & --0.15  &  --0.66  & & --1.65  & --0.10  &  --0.72  & &  --0.90  &  --0.14  &  --0.83   \nl   
O     & --1.78  &  --0.05  & --1.04  & & --1.07  & --0.08  &  --1.11  & & --1.57  & --0.04  &  --1.18  & &  --0.85  &  --0.09  &  --1.30   \nl   
Ne    & --1.49  &  --0.02  & --1.79  & & --1.11  & --0.04  &  --1.86  & & --1.26  & --0.03  &  --1.91  & &  --0.86  &  --0.07  &  --2.02   \nl   
Na    & --2.28  &  --0.00  & --3.02  & & --2.26  & --0.00  &  --3.08  & & --2.14  & --0.00  &  --2.88  & &  --2.01  &  --0.01  &  --2.99   \nl   
Mg    & --2.71  &  --0.45  & --0.19  & & --2.63  & --0.36  &  --0.25  & & --2.52  & --0.35  &  --0.26  & &  --2.44  &  --0.25  &  --0.36   \nl   
Al    & --3.27  &  --0.18  & --0.58  & & --3.29  & --0.15  &  --0.65  & & --3.05  & --0.13  &  --0.70  & &  --3.03  &  --0.10  &  --0.80   \nl   
Si    & --4.05  &  --0.30  & --0.32  & & --4.04  & --0.24  &  --0.39  & & --3.88  & --0.21  &  --0.42  & &  --3.78  &  --0.16  &  --0.53   \nl   
P     & --3.02  &  --0.34  & --0.29  & & --2.94  & --0.28  &  --0.35  & & --2.67  & --0.26  &  --0.37  & &  --2.56  &  --0.19  &  --0.48   \nl   
S     & --3.48  &  --0.38  & --0.24  & & --3.37  & --0.31  &  --0.30  & & --3.13  & --0.28  &  --0.33  & &  --2.98  &  --0.21  &  --0.43   \nl   
Ar    & --2.42  &  --0.36  & --0.26  & & --1.57  & --0.30  &  --0.32  & & --2.21  & --0.33  &  --0.28  & &  --1.33  &  --0.26  &  --0.40   \nl   
Ca    & --3.08  &  --1.21  & --0.03  & & --3.05  & --1.14  &  --0.03  & & --2.64  & --0.97  &  --0.05  & &  --2.56  &  --0.85  &  --0.07   \nl   
Ti    & --3.74  &  --0.74  & --0.20  & & --3.54  & --0.55  &  --0.25  & & --3.51  & --0.59  &  --0.22  & &  --3.21  &  --0.39  &  --0.31   \nl   
Cr    & --4.04  &  --0.89  & --0.12  & & --3.81  & --0.63  &  --0.17  & & --3.76  & --0.70  &  --0.14  & &  --3.46  &  --0.44  &  --0.23   \nl   
Mn    & --2.32  &  --0.67  & --0.15  & & --2.05  & --0.51  &  --0.20  & & --2.00  & --0.53  &  --0.19  & &  --1.69  &  --0.37  &  --0.28   \nl   
Fe    & --4.03  &  --0.90  & --0.16  & & --3.77  & --0.64  &  --0.21  & & --3.75  & --0.73  &  --0.17  & &  --3.38  &  --0.47  &  --0.26   \nl  
Ni    & --2.56  &  --0.47  & --0.21  & & --2.51  & --0.37  &  --0.27  & & --2.28  & --0.34  &  --0.30  & &  --2.17  &  --0.24  &  --0.41   \nl  
Zn    & --2.57  &  --0.22  & --0.41  & & --2.54  & --0.18  &  --0.48  & & --2.41  & --0.17  &  --0.50  & &  --2.26  &  --0.13  &  --0.61   \nl  
\enddata
\tablenotetext{a}{The ionization fraction of an ion $X^i$ is given by 
	$x(X^i) = N(X^i)/N(X_{\rm Total})$.  We present
	radially--averaged values of $\log x(X^i)$ for the first three
	ionization states of each element. Each model assumes WNM abundances (see Table~2) and
	a stellar input spectrum characterized by $T_{eff}$ = 35,000 K
	and log\,(g) = 4.0.  Results are listed for ionization parameters of
	 log\,$(q)$ = --3.0 and --4.0.}
\end{deluxetable}